\documentclass[12pt,draftclsnofoot, perreview, onecolumn]{IEEEtran}
\usepackage{balance}
\usepackage{color}
\usepackage{url}
\usepackage{threeparttable}
\usepackage{amsfonts}
\usepackage{amsmath}
\usepackage{mathrsfs}
\usepackage{multirow}
\usepackage{amssymb}
\usepackage{mdwmath}
\usepackage{bm}
\usepackage{ifpdf}

\ifCLASSINFOpdf
  \usepackage[pdftex]{graphicx}
  \graphicspath{{../pdf/}{../jpeg/}}
  \DeclareGraphicsExtensions{.pdf,.jpeg,.png}
\else
  \usepackage[dvips]{graphicx}
  \graphicspath{{../eps/}}
  \DeclareGraphicsExtensions{.eps}
\fi

\ifCLASSOPTIONcompsoc
  \usepackage[tight,normalsize,sf,SF]{subfigure}
\else
  \usepackage[tight,footnotesize]{subfigure}
\fi

\usepackage{pict2e}
\makeatletter
\newcommand*{\bigboxplus}{\DOTSB\mathop{\mathpalette\big@boxplus\relax}\slimits@}

\newcommand{\big@boxplus}[2]{%
  \vcenter{%
    \m@th\bigbox@thickness{#1}%
    \sbox\z@{$#1\bigoplus$}%
    \dimen@=\ht\z@ \advance\dimen@\dp\z@
    \hbox{%
      \setlength{\unitlength}{\dimen@}%
      \begin{picture}(1,1)
      \polyline(0.1,0.1)(0.9,0.1)(0.9,0.9)(0.1,0.9)(0.1,0.1)(0.5,0.1)
      \polyline(0.5,0.1)(0.5,0.9)
      \polyline(0.1,0.5)(0.9,0.5)
      \end{picture}%
    }%
  }%
}

\newcommand{\bigbox@thickness}[1]{%
  \ifx#1\displaystyle
    \linethickness{0.2ex}%
  \else
    \ifx#1\textstyle
      \linethickness{0.16ex}%
    \else
      \ifx#1\scriptstyle
        \linethickness{0.12ex}%
      \else
        \linethickness{0.1ex}%
      \fi
    \fi
  \fi
}
\makeatother

\newtheorem{algorithm}{\textbf{Algorithm}}
\newtheorem{example}{\textbf{Example}}

\ifCLASSOPTIONonecolumn
\newcommand{\figwidth}{0.65\textwidth}

\newcommand{\vpscaefigure}{\vspace{0.0cm}}

\fi

\ifCLASSOPTIONtwocolumn
\newcommand{\figwidth}{0.48\textwidth}

\newcommand{\vpscaefigure}{\vspace{0.0cm}}

\fi

\begin{document}

\title{Block Markov Superposition Transmission of BCH Codes with Iterative Erasures-and-Errors Decoders}

\author{Suihua~Cai,
        Nina~Lin,
        and~Xiao~Ma,~\IEEEmembership{Member,~IEEE,}
\thanks{This work was supported by the 973 Program (No. 2012CB316100), the NSF of China (No. 91438101), and the NSF of Guangdong (No. 2016A030308008).}
\thanks{This paper was presented in part at 2017 IEEE International Symposium on Information Theory.}
\thanks{Suihua~Cai and Nina~Lin are with the School of Electronics and Information Technology, Sun Yat-sen University, Guangzhou 510006, China (e-mail: caish5@mail2.sysu.edu.cn, linnina@mail2.sysu.edu.cn).}
\thanks{Xiao~Ma is with the School of Data and Computer Science and Guangdong Key Laboratory of Information Security Technology, Sun Yat-sen University, Guangzhou 510006, China (e-mail: maxiao@mail.sysu.edu.cn).}
}

\maketitle
\begin{abstract}
In this paper, we present the block Markov superposition transmission of BCH~(BMST-BCH) codes, which can be constructed to obtain a very low error floor.
To reduce the implementation complexity, we design a low complexity iterative sliding-window decoding algorithm, in which only binary and/or erasure messages are processed and exchanged between processing units.
The error floor can be predicted by a genie-aided lower bound, while the waterfall performance can be analyzed by the density evolution method.
To evaluate the error floor of the constructed BMST-BCH codes at a very low bit error rate~(BER) region, we propose a fast simulation approach.
Numerical results show that, at a target BER of $10^{-15}$, the hard-decision decoding of the BMST-BCH codes with overhead $25\%$ can achieve a net coding gain~(NCG) of $10.55$~dB.
Furthermore, the soft-decision decoding can yield an NCG of $10.74$~dB.
The construction of BMST-BCH codes is flexible to trade off latency against performance at all overheads of interest and  may find applications in optical transport networks as an attractive~candidate.
\end{abstract}


%

\section{Introduction}
In modern optical transport networks~(OTN), forward error correction~(FEC) plays a key role in ensuring reliable transmission. 
Since the first FEC scheme was introduced into optical communications by Grover in 1988~\cite{Grover88}, the FEC schemes used in OTN are classified in three generations.

The first-generation FEC schemes use conventional hard-decision block codes, such as Hamming codes, Reed-Solomon~(RS) codes and Bose-Chaudhuri-Hocquenghem~(BCH) codes.
For example, the RS~[255,239] was recommended for long-haul optical transmission as defined by ITU-T~G$.709$ standard~\cite{G709}.
The performance of this code generation is expected to yield a net coding gain~(NCG) of $6$~dB at an output BER of $10^{-12}$, which has been successfully used in the 2.5~G transmission system.

The second-generation FEC schemes also use the hard-decision decoding~(HDD).
The concatenated product code structures are used to obtain high NCG at an output BER of $10^{-15}$.
The RS codes and the BCH codes are often taken as the component codes, e.g., RS-CSOC~(convolutional self-orthogonal code) having an NCG of $8.3$~dB~\cite{Seki02}, RS-BCH product code having an NCG of $8.67$~dB, and BCH-BCH product code having an NCG of $9.24$~dB~\cite{G9751}.

The third-generation FEC schemes use the technique of soft-decision decoding~(SDD), of which the block turbo codes~(BTC) and the LDPC codes are the two competing classes~\cite{Benedetto05}.
Although the error correction capability can be further improved by a soft-decision FEC, it is believed that the hard-decision FEC is the most feasible to implement as 40~G and 100~G transmission systems~\cite{Chang10}, due to the high processing complexity of  codes and the cost of the required Analog-to-Digital~(A/D) converters.

Recently, new hard-decision FEC schemes are presented, such as the staircase codes~\cite{Smith12}, the continuous-interleaved codes~(CI-BCH)~\cite{Scholten10} and the braided BCH codes~\cite{Jian13}.
These codes can be considered as a combination of concatenated product codes and convolutional structured codes~\cite{Leven14}.
In particular, the staircase code, an elegant extension from a block BCH-BCH product code to its convolutional counterpart, achieves an NCG of $9.41$~dB.
However, the design of staircase codes is inflexible to adapt the change of frame sizes and overheads.
Furthermore, the construction of staircase codes usually requires a large amount of calculation for searching good code parameters either by the extensive software simulations~\cite{Zhang14} or by density evolution (DE) analysis~\cite{Hager15}.

In this paper, we use the block Markov superposition transmission~(BMST) system~\cite{Ma13} to construct high-performing FEC.
The performance of the BMST codes can be predicted by a simple genie-aided bound~\cite{Ma15}, and good BMST codes can be constructed following the general procedure given in~\cite{Liang14}.
To construct FEC with high rates and achieving very low error probabilities, we use the (possibly shortened)~BCH codes as the component codes.
Different from BMST systems with other types of component codes, it is difficult for BCH decoders to obtain and handle truly soft information such as log-likelihood ratios.
Instead, we introduce an erasure message for decoding failure (not as implemented in~\cite{Bosco03}) and presented a soft iterative decoding algorithm~\cite{Lin17} in which only binary and/or erasure messages are processed and exchanged between proceeding units.
In this paper, we further improve the NCG performance with slight increase in complexity by making a three-level decision on the channel output.

The main contributions of this paper are as follows.
\begin{enumerate}
\item A new type of convolutional BCH codes is proposed, in which short BCH codes are embedded into the BMST system, resulting in BMST-BCH codes.
    We show by simulation that, by adjusting the Cartesian product order of the component BCH codes and the encoding memory of the BMST system, the construction is flexible to trade off latency against performance.
\item A low complexity soft iterative sliding-window decoding~(SWD) algorithm is presented.
    The algorithm is both hard-decision and soft-decision capable, in which only binary and/or erasure messages are processed and exchanged between processing units.
    Simulation result shows that, BMST-BCH codes with HDD can achieve comparable NCG performance against the staircase codes, while the BMST-BCH codes with SDD perform even better.
\item Genie-aided lower bound and the DE analysis for the BMST-BCH codes are presented, which are used for predicting and evaluating the error correction capability at a very low BER. To reduce time consumption on the analysis, a fast simulation for BCH codes is also proposed.
\end{enumerate}

The rest of this paper is organized as follows.
The encoding and decoding scheme of the BMST-BCH codes are introduced in Section II.
The genie-aided lower bound and the DE analysis are presented in Section III, where also introduced is the fast simulation method for the BCH codes.
Numerical results are presented in Section IV.
Finally, Section V summarizes this paper.

\section{Encoding and Decoding of BMST-BCH Systems}

\subsection{Channel Model}\label{sec:Channel}
In this paper, we consider binary phase-shift keying~(BPSK) signalling over additive white Gaussian noise~(AWGN) channels. The bits are modulated to $\mathcal{X}=\{+1,-1\}$. Let $Z$ denotes the received signal with
$Z=X+W$, where $W$ is a normal distributed random variable with mean zero.
For ease of implementation, we make a soft decision with three levels.
\begin{equation}
Y=\left\{\begin{array}{ll}
        0, & Z>T, \\
        e, & -T\leqslant Z\leqslant T,\\
        1, & Z<-T,
        \end{array}\right.
\end{equation}
where $T$ is a non-negative threshold to be determined, $Y$ represents the soft decision output and the symbol $e$ represents a bit erasure.
Hence, the channel model is equivalent to a binary symmetric erasure channels~(BSEC) with input alphabet $\mathbb{F}_2=\{0,1\}$ and output alphabet $\{0,1,e\}$.
Such a channel can be characterized by a probability vector $(p_0,p_1,p_e)$, where $p_0$ is the probability of a bit being correctly transmitted, $p_1$ is the probability of bit error and $p_e$ is the probability of bit erasure.
These parameters can be calculated as
\begin{align}
\label{equ:BSEC0}
p_0=&1-Q(\frac{1-T}{\sigma}),\\
p_1=&Q(\frac{1+T}{\sigma}),\\
p_e=&Q(\frac{1-T}{\sigma})-Q(\frac{1+T}{\sigma}),
\label{equ:BSEC1}
\end{align}
where $\sigma$ is the standard deviation of the Gaussian noise, and the Q-function is defined as
\begin{equation}
Q(x)\triangleq\int_x^{\infty}\frac{1}{\sqrt{2\pi}}\exp(-\frac{t^2}{2})\mathrm{d}t=\frac{1}{2}\mathrm{erfc}(\frac{x}{\sqrt{2}}).
\end{equation}

Obviously, different threshold $T$ results in different BSEC.
In particular, in the case when $T=0$, HDD is implemented and the channel is reduced to a binary symmetric channel~(BSC) with $p_e=0$ and $p_1=Q\left(\frac{1}{\sigma}\right)$.

It is not difficult to imagine that extra coding gain can be obtained by selecting a proper value of the threshold.
In this paper, following~\cite{Ma02}, an SNR-dependent threshold $T$ is used, which is calculated numerically according to the mutual information~(MI) criterion.
For an independent and uniformly distributed information source, the MI of BSEC can be calculated as
\begin{equation}
I(X;Y)=\!p_0\log p_0\!+\! p_1\log p_1\!-\! (1\!-\!p_e)\log\frac{1\!-\!p_e}{2}.
\end{equation}
For a given SNR, the parameters $p_0$, $p_1$ and  $p_e$ are functions of $T$. Hence, the threshold $T$  can be optimized by
\begin{equation}
T^*=\arg\max_{T\geqslant 0}I(X;Y).
\end{equation}
\subsection{Encoding of BMST-BCH Systems}
Let BCH~$[n,k, d_{\min}]$ be a binary systematic BCH code, where $n$ is the code length, $k$ is the code dimension and $d_{\min}$ is the designed distance.
By definition, the code rate is $R\triangleq\frac{k}{n}$, and the overhead is calculated as $\frac{n-k}{k}$.
An erasures-and-errors decoder is assumed to correct both erasures and errors.
In the case when the number $i$ of occurrences of errors and the number $j$ of occurrences of erasures satisfy $2i+j < d_{\min}$, the erasures-and-errors decoder outputs the transmitted codeword and declares a decoding success.
In the other case, the erasures-and-errors decoder may declare a decoding success, corresponding to a miscorrection~(an undetected error), or a decoding failure, corresponding to a detected error.

\begin{figure}[t]
\centering
\includegraphics[width=\figwidth]{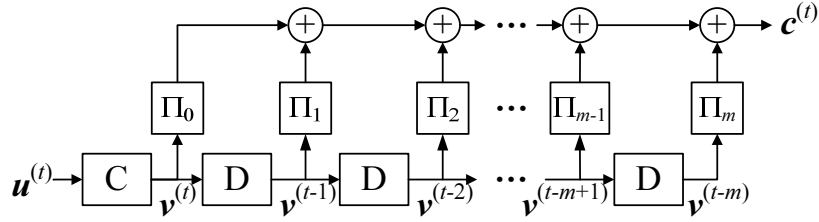}
\caption{The encoding diagram of a BMST-BCH system with memory $m$.}
\label{fig:SystemModel}\vpscaefigure
\end{figure}

Given a positive integer $B$, we take BCH~$[n,k,d_{\min}]^B$, the $B$-fold Cartesian product of BCH~$[n,k,d_{\min}]$, as the basic code to build the BMST system.
Let $\bm{u}^{(0)}$, $\bm{u}^{(1)}$, $\dots$, $\bm{u}^{(L-1)}$ be $L$ blocks of data to be transmitted, where $\bm{u}^{(t)} \in \mathbb{F}_2^{kB}$.
The encoding algorithm of BMST-BCH with memory $m$ is described in Algorithm~\ref{alg:Encode}, see Fig. \ref{fig:SystemModel} for reference.

\begin{algorithm}{Encoding of BMST-BCH}\label{alg:Encode}
\begin{itemize}
\item {\em Initialization}: For $t<0$, set $\bm{v}^{(t)}=\bm{0}\in \mathbb{F}_2^{nB}$.
\item {\em Iteration}: For $t=0,1,\dots,L-1$,
\begin{enumerate}
\item[1)] Encode $\bm{u}^{(t)}$ into $\bm{v}^{(t)} \in \mathbb{F}_2^{nB}$ by performing separately $B$ times the encoding algorithm of the component code BCH~$[n,k,d_{\min}]$;
\item[2)] For $0\leqslant i \leqslant m$, interleave $\bm{v}^{(t-i)}$ into $\bm{w}^{(t,i)}$ by the $i$-th interleaver $\bm{\Pi}_i$~(randomly generated but fixed);
\item[3)] Compute $\bm{c}^{(t)}=\sum_{i=0}^m\bm{w}^{(t,i)}$, which is taken as the $t$-th block of transmission.
\end{enumerate}
\item  {\em Termination}: For $t=L,L+1,\dots,L+m-1$, set $\bm{u}^{(t)}=\bm{0}\in \mathbb{F}_2^{kB}$ and compute $\bm{c}^{(t)}$ following Step {\em Iteration}.
\end{itemize}
\end{algorithm}
{\bf Remark.} The code rate is $\frac{kL}{n(L+m)}$, which is slightly less than that of the basic code.
However, the rate loss is negligible for the case when $m\ll L$.

\subsection{Sliding-window Decoding Algorithm}
\begin{figure}[t]
\centering
\includegraphics[angle=270,width=\figwidth]{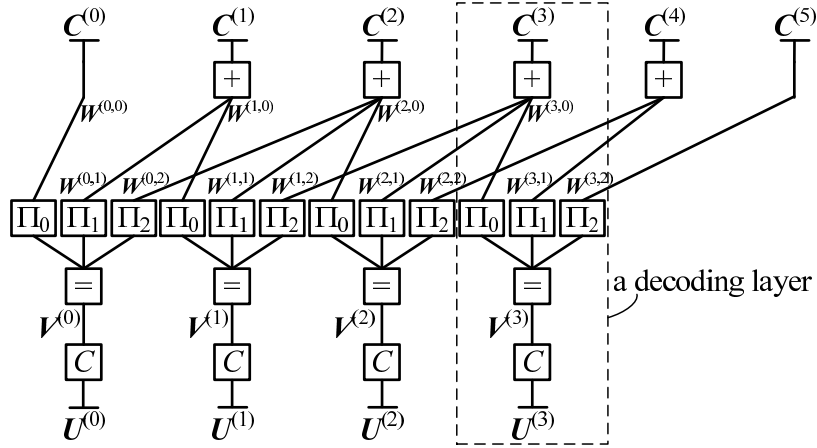}
\caption{The normal graph of a BMST-BCH system with $L=4$ and $m=2$.}
\label{fig:NormalGraph}\vpscaefigure
\end{figure}

As a special class of BMST codes, BMST-BCH codes can be represented by a normal graph~\cite{Ma15}, where edges represent variables and vertices (nodes) represent constraints.
The iterative SWD algorithm with decoding delay $d$ can be described as a message processing/passing algorithm over a subgraph containing $d+1$ layers, where each layer consists of a node of type \fbox{C}, a node of type \fbox{=}, $m+1$ nodes of type \fbox{$\Pi$}, and a node of type \fbox{+}.
See Fig. \ref{fig:NormalGraph} for reference.
The decoding algorithm starts from nodes of type \fbox{+}, which are connected to channels and receive messages $\bm{z}^{(t)}$, the noisy versions of $\bm{c}^{(t)}$.


\begin{itemize}
\item {\bf Node} \fbox{+}: This type of nodes has $m+2$ edges, where $m+1$ edges are connected to nodes of \fbox{$\Pi$}, which carry the messages of $\bm{w}$'s, and a half-edge connected to the channel output.
    To make this simpler, we use the notation for the $i$-th edge with the input messages $\bm{x}_i\in \{0,1,e\}^{nB}$ and the output messages $\bm{z}_i\in \{0,1,e\}^{nB}$, for $0\leqslant i\leqslant m+1$, without distinguishing the two types of edges.
    We define a binary operator $\boxplus$ on $\{0,1,e\}$ as
    \begin{equation}
    a \boxplus b =\left\{\begin{array}{ll}a \oplus b,& a \ne e \text{ and } b \ne e,\\ e,&{\rm otherwise}, \end{array}\right.
    \label{equ:BoxPlus}
    \end{equation}
    where $\oplus$ represents the addition on $\mathbb{F}_2$.
    Given the input messages $\bm{x}_{i}$, $0 \leqslant i \leqslant m + 1$, the output messages $\bm{z}_{i}$ are calculated by summing up the corresponding input messages from the other edges,
    \begin{equation}
    z_{i,\ell}=\bigboxplus_{j\ne i}x_{j,\ell}, \mathrm{~for~}0 \leqslant \ell \leqslant nB-1.
    \end{equation}
\item {\bf Node} \fbox{$\Pi_i$}: The node \fbox{$\Pi_i$} represents the $i$-th interleaver, which interleaves or de-interleaves the input messages.
\item {\bf Node} \fbox{=}:
    With a slight abuse of notation, we still use $\bm{x}_i$ and $\bm{z}_i$ to denote the input messages and output messages associated with the $i$-th edge, for $0\leqslant i \leqslant m+1$.
    For convenience, the edge connected to node \fbox{C} is numbered by $i=0$.
    For \fbox{=} $\rightarrow$ \fbox{C}, the $\ell$-th component of the output message $\bm{z}_0$ is calculated by voting as
    \begin{equation}
        \!\!\!z_{0,\ell} \!= \! \left\{\begin{array}{ll}
        \!\!0, & \vert\!\{i\ne0 \vert x_{i,\ell} = 0\}\! \vert \!>\! \vert\!\{i\ne 0 \vert x_{i,\ell} = 1\}\! \vert, \\
        \!\!1, & \vert\!\{i\ne0 \vert x_{i,\ell} = 0\}\! \vert \!<\! \vert\!\{i\ne 0 \vert x_{i,\ell} = 1\}\! \vert,\\
        \!\!e, & \textrm{otherwise}.
        \end{array}\right.
    \end{equation}
    For \fbox{$=$} $\rightarrow$ \fbox{$\Pi$}, the output messages $z_{i,\ell}$, $1 \leqslant i \leqslant m + 1$, $0 \leqslant \ell \leqslant nB-1$, are set to be $x_{0,\ell}$ if $x_{0,\ell} \ne e$, or calculated by voting if $x_{0,\ell}= e$.
    That is,
    \begin{equation}
        \!\!\!z_{i,\ell} \!=\! \!\left\{\begin{array}{lll}
        \!\!\!\!x_{0,\ell},\!\!\!\!\!& \!& x_{0,\ell}\! \ne\! e, \\
        \!\!\!\!0,\!\!\!\!\!& \!& x_{0,\ell} \!= \!e \!\textrm{~and~}\! \vert\!\{\!j \!\ne \!i \vert x_{j,\ell} \!= \!0\!\}\! \vert\! > \! \vert\!\{\!j\! \ne \!i \vert x_{j,\ell}\! = \!1\!\}\! \vert,\\
        \!\!\!\!1,\!\!\!\!\!&\!& x_{0,\ell} \!=\! e \!\textrm{~and~}\! \vert\!\{\!j\! \ne\! i \vert x_{j,\ell}\! = \! 0\!\} \!\vert\! <\! \vert\!\{\!j\! \ne\! i \vert x_{j,\ell}\! = \!1\!\}\! \vert,\\
        \!\!\!\!e,\!\!\!\!\!& \!& \textrm{otherwise}.
        \end{array}\right.
    \end{equation}
\item {\bf Node} \fbox{C}: Given the messages from node \fbox{=} are available, the node \fbox{C} performs an erasures-and-errors decoding algorithm to compute the extrinsic messages. Specifically, the decoder outputs a codeword if it is successful or a sequence of erasures $\bm{e}=(e,e,\dots,e)$ otherwise.
\end{itemize}

{\bf Remark.} For the nodes \fbox{+} and \fbox{=} with $m+2$ edges connected, by using the ``partial sum'' technique presented at Section~5.4 of~\cite{Ryan09}, only $3m$ binary operations for vectors are needed for each node processing.

The decoding algorithm is summarized in Algorithm~\ref{alg:Decode} as follows.

\begin{algorithm}{SWD of BMST-BCH}\label{alg:Decode}
\begin{itemize}
\item  {\em Global initialization}: 
    Set a maximum iteration number $I_{\rm max} > 0$. The input messages of half-edges connected to the $t$-th layer are initialized as the corresponding received messages $\bm{y}^{(t)} \in\{0,1,e\}^{nB}$ for $0\leqslant t\leqslant d-1$ and the other edges are initialized as message $e$.
\item  {\em Sliding window}: For $t=0,1,\dots,L-1$,

    \begin{itemize}
        \item[1)] \textbf{\textit{Local initialization}}: Initialize the input messages of half-edge connected to the $(t+d)$-th layer as the received messages $\bm{y}^{(t+d)}$.
        \item[2)] \textbf{\textit{Iteration}}: For $I=1,2,\dots, I_{\rm max}$,
        \begin{enumerate}
            \item[a)] {\em Forward recursion}: For $i=0,1,\dots,d$, the $(t+i)$-th layer performs a message processing algorithm scheduled as
        {\center \fbox{+} $\rightarrow$ \fbox{$\Pi$} $\rightarrow$ \fbox{=} $\rightarrow$ \fbox{C} $\rightarrow$ \fbox{=} $\rightarrow$ \fbox{$\Pi$} $\rightarrow$ \fbox{+}.}
            \item[b)] {\em Backward recursion}: For $i=d,d-1,\dots,0$, the $(t+i)$-th layer performs a message processing algorithm scheduled as
        {\center \fbox{+} $\rightarrow$ \fbox{$\Pi$} $\rightarrow$ \fbox{=} $\rightarrow$ \fbox{C} $\rightarrow$ \fbox{=} $\rightarrow$ \fbox{$\Pi$} $\rightarrow$ \fbox{+}.}
            \item[c)] {\em HDD output}: If a decoding success at node \fbox{C} is declared, the recovered information is output, or if a decoding failure is declared, the systematic part of the messages from \fbox{=} is output directly.
                    In the latter case, each message $e$ in the systematic part is replaced independently and uniformly at random by either $0$ or $1$.
                    If the output messages at node \fbox{C} stay unchanged for some iteration $I>1$, exit the iteration.
        \end{enumerate}
        \item[3)] \textbf{\textit{Cancelation}}: Remove the effect of $\hat{\bm{v}}^{(t)}$ on all layers by updating the input messages as
            \begin{equation}
                \bm{y}^{(t + i)} \leftarrow \bm{y}^{(t + i)} \boxplus \hat{\bm{w}}^{(t, i)},
            \end{equation}
            for  $i = 1, 2, \cdots, m$. Here we use the notation with a hat sign~(~$\hat{}$~) to denote the corresponding estimated messages according to the decoder output.
    \end{itemize}
\end{itemize}
\end{algorithm}

\subsection{Complexity Analysis}

We will show with numerical examples that both the encoding memory and the decoding delay can be chosen with a small amount to fulfill the performance requirement by choosing properly the component BCH code.
For the encoding scheme, as seen in Fig.~\ref{alg:Encode}, the binary messages are processed by a basic code encoder, $m$ registers, $m+1$ interleavers and $m$ summators.
Thus, the encoding complexity has the same scale as the encoding complexity of the component BCH code.

For the decoding scheme, as seen in Fig.~\ref{alg:Decode}, only simple operations are processed at nodes \fbox{+} and \fbox{=}.
Consequently, the computational complexity of message processing at nodes \fbox{+} and \fbox{=} of degree $m+2$ is much less than the decoding complexity of the component BCH code, roughly $\mathcal{O}(n^2)$~\cite{Bajoga73}.
On the other hand, the $B$-folds of the component BCH cods are independent of each other, in which parallel operation can be used in the encoding/decoding schemes for the basic codes.
As a result, the encoding/decoding complexity of the BMST-BCH codes, when averaged over the data length, is almost linear with that of the component BCH code.

\section{Performance Analysis}
To analyze the BER performance of BMST-BCH codes, we present both the genie-aided lower bound and the DE analysis.
As we will show by simulation, the DE analysis not only verifies the achievability of the genie-aided lower bound, but also confirms the near-optimality of the choice of the decoding delay $d=2m$.
To overcome the difficulty caused by the conventional time-consuming simulations, we propose a fast simulation approach as an essential tool to evaluate the lower bounds and the DE thresholds.

\subsection{Fast Simulation Approach for BCH Codes}
It is typically time-consuming to simulate the performance of a code in the extremely low error region.
To overcome this problem, we write the BER as
\begin{equation}
{\rm BER}=\sum_{i,j}P_{i,j}{\rm BER}_{i,j},
\label{equ:ProSim}
\end{equation}
where $P_{i,j}$ is the probability that the channel introduces $i$ errors and $j$ erasures to the transmitted BCH codeword of length $n$, and ${\rm BER}_{i,j}$ is the conditional BER given that these errors and erasures are uniformly distributed.
Since the bit errors occurrence are considered independent and identically distributed~(i.i.d.), it is not difficult to check that $P_{i,j}$ has a closed form as
\begin{equation}
P_{i,j}=\binom{n}{i}\binom{n-i}{j}p_1^ip_e^jp_0^{n-i-j}\triangleq f_{i,j}(p_0,p_1,p_e).
\label{equ:Proij}
\end{equation}
Recalling that the erasures are replaced by random bits in the case of decoding failure, the conditional error probability ${\rm BER}_{i,j}$ can be written as
\begin{equation}
{\rm BER}_{i, j} =  \mu_{i,j}  +\lambda_{i,j}(\frac{i}{n} + \frac{j}{2n}),
\label{equ:ProBER}
\end{equation}
where $\mu_{i, j}$ is the bit error rate associated with the successful decoding (i.e., the ratio between the number of bit errors of successful decoding and the total number of information bits), and $\lambda_{i,j}$ is the probability of decoding failure.
It can be checked that
\begin{align}
&\mu_{i,j}=0,~\lambda_{i,j}=0,~~{\rm ~for~}2i+j<d_{\min}.\\
&\mu_{i,j}=0,~\lambda_{i,j}=1,~~{\rm ~for~}j\geqslant d_{\min}.
\end{align}
For the remaining cases when $2i+j \geqslant d_{\min}$ and $j<d_{\min}$, $\mu_{i, j}$ and $\lambda_{i,j}$ can be estimated efficiently with Monte Carlo simulations (described in Algorithm~\ref{alg:Simu}) by imposing uniformly at random $i$ errors and $j$ erasures on a codeword.
\begin{algorithm}{Monte Carlo Simulation of $\mu_{i,j}$ and $\lambda_{i,j}$ for BCH~$[n,k,d_{\min}]$ for given $(i,j)\in\{2i+j \geqslant d_{\min},~j<d_{\min}\}$}\label{alg:Simu}
\begin{itemize}
\item  {\em Initialization}: Set a sampling size $S$. Initialize the counters $n_1=0,~n_2=0$.
\item  {\em Iteration}: For $s=1,2,\dots,S$,
    \begin{itemize}
        \item[1)] Generate a random vector $\bm{v}\in\{0,1,e\}^n$, consisting of $n-i-j$ components of value $0$, $i$ components of value $1$ and $j$ components of value $e$ in a totally random order.
        \item[2)] Decode $\bm{v}$ by the erasures-and-errors decoding algorithm.
        \begin{enumerate}
            \item[a)] For the case of decoding success, increase the counter $n_1$ by the Hamming weight of the output codeword.
            \item[b)] For the case decoding failure, increase the counter $n_2$ by one.
        \end{enumerate}
    \end{itemize}
\item {\em Output}: $\mu_{i,j}\approx\frac{n_1}{nS}$, $\lambda_{i,j}\approx\frac{n_2}{S}$.
\end{itemize}
\end{algorithm}
{\bf Remark.} It is worth pointing out that, by definition, $\mu_{i,j}$, $\lambda_{i,j}$ and ${\rm BER}_{i,j}$ depend only on the code structure and are independent of the channel parameters.
We assume in the sequel that these values are available.

\subsection{Genie-aided Lower Bound}
The genie-aided lower bound is derived by assuming all but one layers~(sub-blocks) are perfectly known at the decoder.
This is equivalent to assuming that a BCH codeword is transmitted $m + 1$ times over the BSEC~$(p_0, p_1, p_e)$.
Hence, after messages processing at node \fbox{$=$}, the BCH decoder is faced with a BSEC~$(\tilde{p}_0, \tilde{p}_1, \tilde{p}_e)$, where
\begin{equation}
\tilde{p}_0=\sum_{i=1}^{m+1}\sum_{j=0}^{i-1}\binom{m+1}{i}\binom{m+1-i}{j}p_0^ip_1^jp_e^{m+1-i-j},
\label{equ:tilde0}
\end{equation}
\begin{equation}
\tilde{p}_1=\sum_{i=1}^{m+1}\sum_{j=0}^{i-1}\binom{m+1}{i}\binom{m+1-i}{j}p_0^jp_1^ip_e^{m+1-i-j},
\label{equ:tilde1}
\end{equation}
and
\begin{equation}
\tilde{p}_e=\sum_{i=0}^{\lfloor(m+1)/2\rfloor}\binom{m+1}{2i}\binom{2i}{i}p_0^{i}p_1^{i}p_e^{m+1-2i}.
\label{equ:tildee}
\end{equation}
The lower bound of the BMST-BCH system is then calculated according to~(\ref{equ:ProSim}) with $P_{i,j}=f_{i,j}(\tilde{p}_0,\tilde{p}_1,\tilde{p}_e)$ as expressed in~(\ref{equ:Proij}).

\subsection{Density Evolution}
DE is a tool to predict the iterative decoding performance of asymptotically long codes in the waterfall region by simulation on the decoding process under a cycle-free graph assumption by tracking how the probability densities of the exchanged messages evolve with iterations.
The DE analysis is different from that presented in~\cite{Jian12} in the following two aspects.
First, an erasure message is allowed and a probability vector $(\alpha,\beta,\gamma)$ is used to represent the distribution of the propagating messages in the decoding algorithm, where $\alpha$ is the probability of a bit being correct, $\beta$ is the probability of bit flipping error and $\gamma$ is the probability of bit erasure.
Second, we will not ignore the case of miscorrection and do not assume the availability of the weight spectrum.
For clarity, we take $m=2$ as an example to describe the DE analysis algorithm as follows, where a target BER is preset. For a general case with encoding memory $m$, see Appendix~\ref{sec:de_g}.

\begin{algorithm}{DE analysis for SWD of BMST-BCH code with $m=2$}\label{alg:DE}
\begin{itemize}
\item  {\em Initialization}: Set a maximum number of iterations $I_{\max}$. All messages over the half-edges~(connected to the channel) are initialized as $(p_0,p_1,p_e)$, and all the messages over the edges are initialized as $(0,0,1)$.
\item  {\em Sliding window}: For each window position, the $d+1$ decoding layers perform iteratively~(see below) message processing layer-by-layer according to the schedule,
{\center~~~~~~~~~~~~~~ \fbox{+} $\rightarrow$  \fbox{=} $\rightarrow$ \fbox{C} $\rightarrow$ \fbox{=} $\rightarrow$ \fbox{+}.}\\
\item[] {\textbf{\textit{Iteration}}:} For $I=1,2,\dots,I_{\max}$,
\begin{enumerate}
\item[1)] {\bf Node} \fbox{+}: At most 4 edges are connected to a node \fbox{+} in total. For each edge, given the input messages $\{(\alpha_{\ell},\beta_{\ell}, \gamma_{\ell}),{\ell}=0,1,2\}$ from the other edges, the extrinsic output messages can be calculated as
\begin{align}
\!\!\!\!\!\!\!\!\!\alpha_{\rm out}&=\alpha_0\alpha_1\alpha_2+\alpha_0\beta_1\beta_2+\alpha_1\beta_2\beta_0+\alpha_2\beta_0\beta_1,\\
\!\!\!\!\!\!\!\!\!\gamma_{\rm out}&=1-(1-\gamma_0)(1-\gamma_1)(1-\gamma_2),\\
\!\!\!\!\!\!\!\!\!\beta_{\rm out}&=1-\alpha_{\rm out}-\gamma_{\rm out}.
\end{align}
\item[2)] {\bf Node} \fbox{=}: From \fbox{=} to \fbox{C}, given the input messages $\{(\alpha_{\ell},\beta_{\ell}, \gamma_{\ell}),{\ell}=0,1,2\}$ from the other edges, the extrinsic  output messages can be calculated as
\begin{align}
\alpha_{\rm out}=&\alpha_0\gamma_1\gamma_2+\alpha_1\gamma_2\gamma_0+\alpha_2\gamma_0\gamma_1\nonumber\\ &\!\!+\alpha_0\alpha_1+\alpha_1\alpha_2+\alpha_2\alpha_0-2\alpha_0\alpha_1\alpha_2,\\
\beta_{\rm out}=&\beta_0\gamma_1\gamma_2+\beta_1\gamma_2\gamma_0+\beta_2\gamma_0\gamma_1\nonumber\\ &\!\!+\beta_0\beta_1+\beta_1\beta_2+\beta_2\beta_0-2\beta_0\beta_1\beta_2,\\
\gamma_{\rm out}=&1-\alpha_{\rm out}-\beta_{\rm out}.
\end{align}
From \fbox{=} to \fbox{+}, given the input messages $\{(\alpha_{\ell},\beta_{\ell}, \gamma_{\ell}),{\ell}=0,1,2\}$ from the other edges (label the input message from node \fbox{C} with index $0$), the extrinsic  output messages are given by
\begin{align}
\alpha_{\rm out}&= \alpha_0+\gamma_0(\alpha_1\gamma_2+\alpha_2\gamma_1+\alpha_1\alpha_2),\\
\beta_{\rm out}&= \beta_0+\gamma_0(\beta_1\gamma_2+\beta_2\gamma_1+\beta_1\beta_2),\\
\gamma_{\rm out}&= 1-\alpha_{\rm out}-\beta_{\rm out}.
\end{align}
\item[3)]{\bf Node} \fbox{C}: Given the input messages $(\alpha,\beta,\gamma)$ from node \fbox{=}, the extrinsic output messages can be calculated as~(\ref{equ:DEBER0}) - (\ref{equ:DEBER1}).
\end{enumerate}
After $I_{\rm max}$ iterations, if the estimated BER is no greater than the target BER, a local decoding success is declared, the window position is shifted, and decoding continues.
Otherwise, a decoding failure is declared and the window decoding terminates.
\end{itemize}

A complete decoding success is declared for a target BER under an initial channel parameter $(p_0,p_1,p_e)$ if and only if all $L$ layers declare decoding success.
A threshold for a target BER is roughly defined as the minimum SNR that guarantees a complete decoding success, which can be found numerically by Algorithm \ref{alg:DE}.
\end{algorithm}

\section{Numerical Results}

All the simulations in this section are conducted by assuming BPSK modulation and AWGN channels.
As mentioned in Section II, for HDD, the channel is modeled as a BSC, equivalent to a BSEC with $p_1=Q(\sqrt{2E_bR/N_0})$, $p_e=0$ and $p_0=1-p_1$.
In this setting of HDD, the external messages received by the SWD are binary only, while the internal messages processed and exchanged by the SWD can be erasures in the case of decoding failure at some node \fbox{C}.
For SDD, the channel is equivalent to a  BSEC  as described in (\ref{equ:BSEC0})-(\ref{equ:BSEC1}).


%
\subsection{Effectiveness of Proposed Evaluations}

In this subsection, a toy example is simulated to show the effectiveness of the proposed evaluation methods.
For this end, we take $10^{-3}\sim 10^{-5}$ as target BERs, at which the fast simulation can be compared with the traditional simulation since both of them are implementable with a reasonable computational resource.

\begin{figure}[t]
\centering
\includegraphics[width=\figwidth]{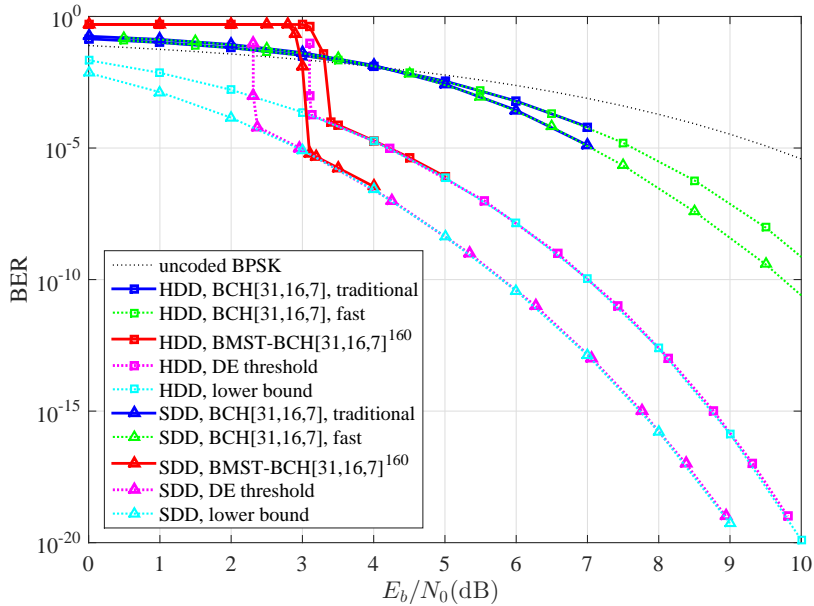}
\caption{Performance of the BMST-BCH system with BCH~$[31, 16, 7]^{160}$ as the basic code.
The system encodes with the encoding memory $m = 2$ and decodes with a decoding delay $d= 4$ and a maximum iteration $I_{\rm max} = 15$.}
\label{fig:Validation}
\end{figure}
\begin{example}\label{exp:Validation}
Consider the BMST-BCH~$[31,16,7]^{160}$ code with encoding memory $m = 2$ and decoding delay $d=4$.
The simulation results are shown in Fig.~\ref{fig:Validation}.
We see that the proposed fast simulation approach matches well with the traditional simulation approach for both HDD and SDD.
As expected, the SDD of the BMST-BCH code leads to a lower error floor and a better waterfall when compared with HDD.
We also see that the DE threshold coincides with the lower bound in the error floor region, suggesting that the bound is tight.
In addition, we observe that the proposed simulation approach can evaluate efficiently the performance around BER of $10^{-15}$.
\end{example}

\subsection{Impact of Parameters on Performance}

In this subsection, we show by numerical results~(including simulation, DE analysis and genie-aided lower bound) how the performance is affected by the parameters such as
the encoding memory $m$, the decoding delay $d$, the Cartesian product order $B$, and
the component BCH code.
As a general methodology, impacts on performance are investigated by varying some parameters while fixing others.

\begin{figure}[t]
\centering
\includegraphics[width=\figwidth]{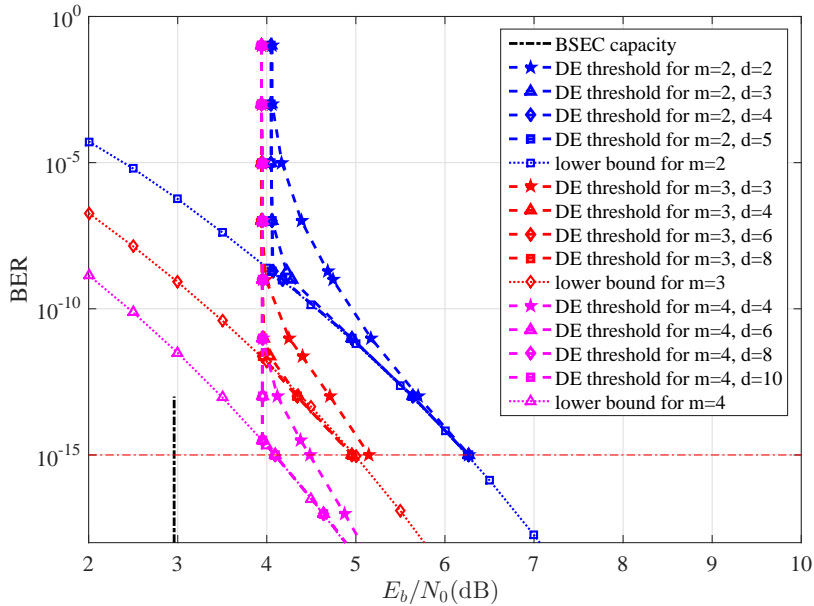}
\caption{DE analysis of the BMST-BCH system with BCH~$[126,105,7]$ as the component code.
The system encodes with the encoding memory $m = 2,3,4$ and decodes~(SDD) with different decoding delays.}
\label{fig:d}
\end{figure}
\begin{example}[Encoding Memory and Decoding Delay]\label{exp:d}
In this example, the BMST-BCH systems are constructed by using BCH~$[126, 105, 7]$ as the component code with varied encoding memory and decoding delay.
The genie-aided lower bounds and the DE analysis of SDD are shown in Fig.~\ref{fig:d}.
The curves have similar behavior to other BMST systems~\cite{Huang16}.
With encoding memory fixed, the DE performance improves as the decoding delay $d$ increases.
However, when $d \geqslant 2m$, the performance improvement is getting saturated as imposed by the genie-aided lower bound, suggesting that the SWD algorithm with decoding delay $d = 2m$ is nearly optimal.
Therefore, we set $d=2m$ in the sequel.
We also see that, as predicted by the genie-aided lower bound, the error floor can be lowered down by increasing the encoding memory.
\end{example}

\begin{figure}[t]
\centering
\includegraphics[width=\figwidth]{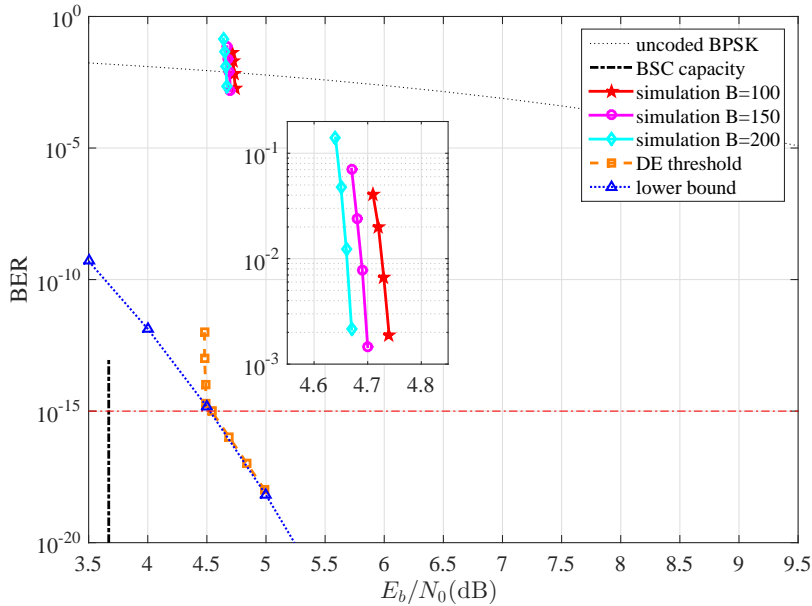}
\caption{Performance of the BMST-BCH system with BCH~$[660,550,23]$ as the component code.
The system encodes with the encoding memory $m = 2$ and decodes~(HDD) with a decoding delay $d = 4$ and a maximum iteration $I_{\rm max} = 15$.}
\label{fig:B}
\end{figure}
\begin{example}[Cartesian Product Order]\label{exp:B}
In this example, we take BCH~$[660, 550, 23]$ as the component code and fix the encoding memory as two.
The simulation results are shown in Fig.~\ref{fig:B}.
We find that the curves are so steep in the waterfall region that we are not able to locate within a reasonable time the required $E_b/N_0$ even for the BER of $10^{-5}$.
However, both the genie-aided lower bound and the DE analysis indicate an NCG\footnote{Notice that the required $E_b/N_0$ for the uncoded system at the BER of $10^{-15}$ is $15.0$~dB.} of $10.48$~dB.
In the finite length regime with $B = 100$, the NCG is reduced to around $10.25$~dB, which is slightly less than $10.41$~dB, the NCG of staircase code given in~\cite{Zhang14} with a similar latency.
This minor degradation of performance, however, is rewarded with an easy way to make trade-offs between latency and performance.
That is, the performance can be improved further to approach the DE threshold by simply increasing $B$.
\end{example}


\begin{figure}[t]
\centering
\includegraphics[width=\figwidth]{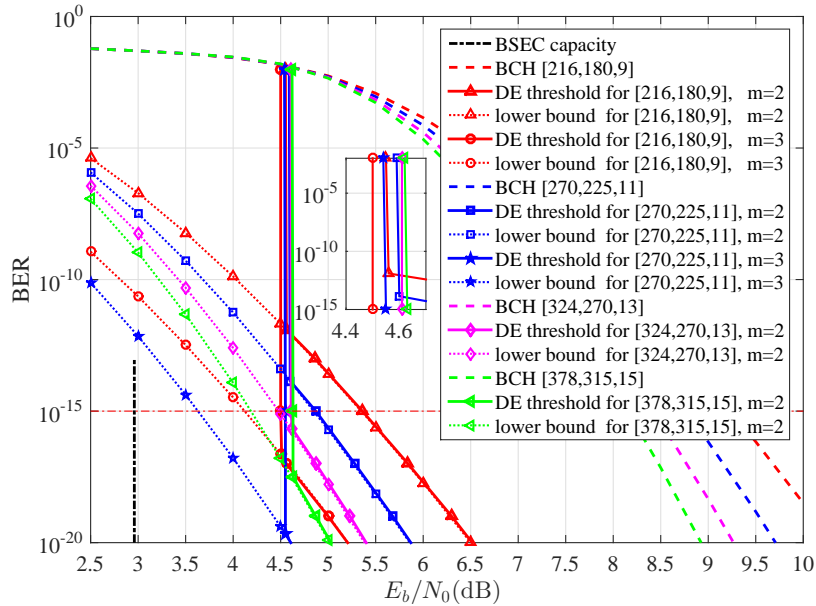}
\caption{DE analysis of the BMST-BCH system with shortened BCH codes defined over GF($2^9$) as the component codes.
The system encodes with the encoding memory $m = 2,3$ and decodes~(SDD) with decoding delay $d = 2m$.}
\label{fig:t}
\end{figure}
\begin{example}[Designed Distance of Component BCH Code]\label{exp:t}
In this example, BCH codes defined over GF($2^9$) are shortened to adapt the BMST systems with an overhead of 20\%.
The simulation results are shown in Fig.~\ref{fig:t}, where the component codes are shortened BCH codes from $[511,475,9]$, $[511,466,11]$, $[511,457,13]$ and $[511,448,15]$.
For $m=2$, we see that the lower bound shifts to the left as the designed distance increases, which is consistent with the BCH code performance.
From the genie-aided lower bound, we see that encoding memory $m=3$ is preferred for $[216,180,9]$ and $[270,180,11]$ to achieve the target BER of $10^{-15}$.
We also see that, with encoding memory fixed, the code has a slightly better DE threshold as the designed distance decreases.
\end{example}

\begin{figure}[t]
\centering
\includegraphics[width=\figwidth]{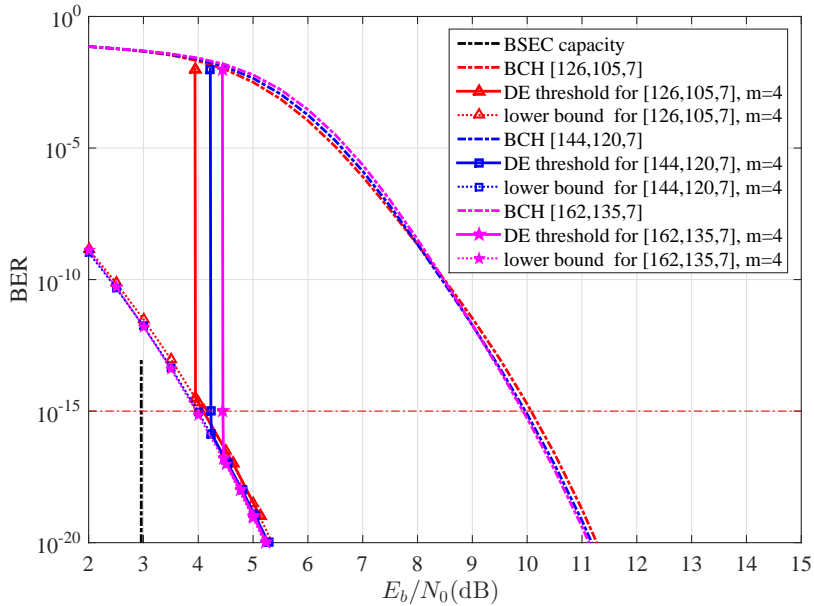}
\caption{DE analysis of the BMST-BCH system using the shortened BCH codes defined over different fields as the component codes.
The system encodes with the encoding memory $m = 4$ and decodes~(SDD) with a decoding delay $d = 8$.}
\label{fig:q}
\end{figure}
\begin{example}[Field Size of Component BCH Code]\label{exp:q}
In this example, BCH codes defined over different fields but with the same designed distance of $7$ are shortened to adapt the BMST systems with an overhead of 20\%.
The simulation results are shown in Fig.~\ref{fig:q}, where the component codes are shortened BCH codes from [127,106,7], [255,231,7] and [511,484,7].
We see that these three systems have similar lower bounds but non-negligible different waterfall performance.
The component code over a larger Galois field results in a slightly better lower bound but a worse DE threshold.
\end{example}

\subsection{Comparison with Staircase Codes}

Compared with the staircase codes, the BMST-BCH codes are easily configured and efficiently optimized.
Given an overhead and a decoding latency, we have the following procedure to search a good code with an extremely low error floor.
\begin{enumerate}
\item List as candidates all BCH codes~(with moderate code length) that satisfy~(after shortening if necessary) the overhead requirement.
\item For each candidate BCH code, select an encoding memory such that the genie-aided lower bound~(obtainable by the fast simulation) lies below the target BER in the SNR region around one dB away from the Shannon limit.
\item Take as component the BCH code~(along with the selected encoding memory) that has an acceptable waterfall performance~(predicted by DE analysis with fast simulation).
\item Calculate the Cartesian product order under the restriction of the decoding latency.
\end{enumerate}


\begin{figure}[t]
\centering
\includegraphics[width=\figwidth]{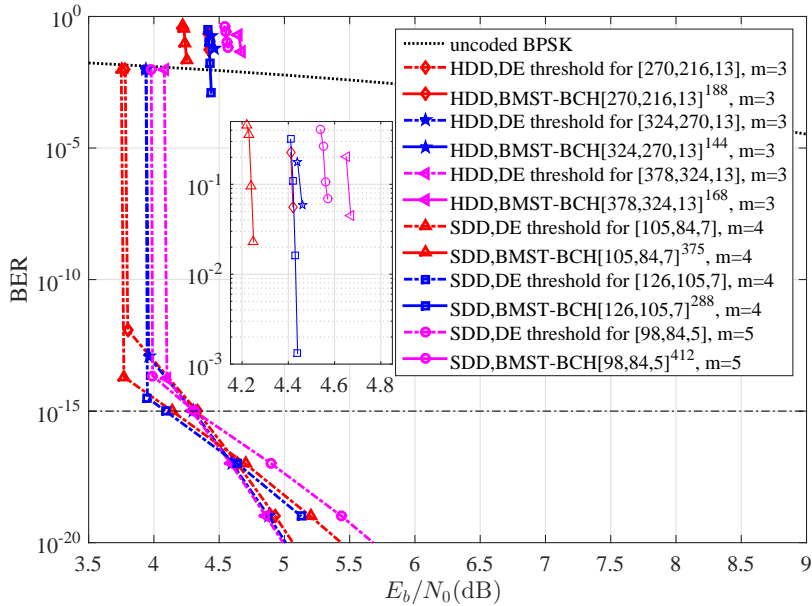}
\caption{Performance of the BMST-BCH systems with overheads of 16.7\%, 20\% and 25\%. }
\label{fig:compare}
\end{figure}

\begin{example}\label{exp:compare}
Following the above construction procedure, we have designed BMST-BCH codes with overheads of 16.7\%, 20\% and 25\% to target the BER of $10^{-15}$ with NCGs as high as possible.
The simulation results are shown in Fig.~\ref{fig:compare}.
Similar to Example~\ref{exp:B}, the curves are so steep in the waterfall region that we are not able to locate within a reasonable time the required $E_b/N_0$ even for the BER of $10^{-5}$.
As evidenced by Example~\ref{exp:Validation}, the NCGs can be estimated by extrapolation based on the simulation results, the lower bounds and the DE analysis.
The NCGs are listed in Table~\ref{tab:NCG}, which also includes the performance of staircase codes reported in~\cite{Zhang14}.
We can see that,with HDD implemented, the BMST-BCH codes with overheads of 16.7\%, 20\% and 25\% have NCGs of $10.32$~dB, $10.53$~dB and $10.55$~dB respectively, which are comparable to those of the staircase codes.
With a similar decoding latency, the SDD results in an NCG improvement within 0.2~dB.
It is worth pointing out that, indicated by the asymptotic performance predicted by the DE analysis, which means, we can further improve the NCGs by increasing the Cartesian product order~$B$.
\end{example}

\begin{table}
  \centering
  \caption{ Performance of BMST-BCH Codes}
  \label{tab:NCG}
  \begin{tabular}{c|c|c|c}
    \hline\hline
    OH$\%$ & code type & latency & NCG(dB)\\

    \hline
            &staircase                  &444 528    &10.32~(HDD)\\
    \cline{2-4}
    16.7    &$[378,324,13]^{168},m=3$   &444 528    &10.32~(HDD)\\
    \cline{2-4}
            &$[98,84,5]^{412},m=5$      &444 136    &10.42~(SDD)\\
    \hline
            &staircase                  &326 592    &10.41~(HDD)\\
    \cline{2-4}
    20      &$[324,270,13]^{144},m=3$   &326 592    &10.53~(HDD)\\
    \cline{2-4}
            &$[126,105,7]^{288},m=4$    &326 592    &10.56~(SDD)\\
    \hline
            &staircase                  &354 375    &10.62~(HDD)\\
    \cline{2-4}
    25      &$[270,216,13]^{188},m=3$   &355 320    &10.55~(HDD)\\
    \cline{2-4}
            &$[105,84,7]^{375},m=4$     &354 375    &10.74~(SDD)\\
    \hline
    \hline
  \end{tabular}

\end{table}

\section{Conclusion}
We have proposed a new type of convolutional BCH codes by embedding short BCH codes into the BMST system, resulting in the BMST-BCH codes.
A soft iterative sliding-window decoding algorithm was also proposed and the NCG performance was analyzed by genie-aided lower bound and DE analysis.
The simulation results showed that the BMST-BCH codes can achieve comparable NCG performance against those of the staircase codes at a target BER of~$10^{-15}$.


%

%
\appendices
\section{Density Evolution for a General Encoding Memory}
\label{sec:de_g}
%
The encoding memory is $m$ and a target BER is preset.
Denote the index set $\mathcal{I}=\{0,1,\dots,m\}$, which labels the input edges.
Operations on the nodes for density evolution analysis are as follows.
\begin{itemize}
\item[1)] {\bf Node} \fbox{+}: At most $m+2$ edges are connected to a node \fbox{+} in total. For each edge, given the input messages $\{(\alpha_{\ell},\beta_{\ell}, \gamma_{\ell}),{\ell}\in \mathcal{I}\}$ from the other edges, the extrinsic output messages can be calculated as
\begin{align}
\alpha_{\rm out}&=\sum_{\substack{\mathcal{A}\subseteq \mathcal{I}\\|\mathcal{A}|{\rm ~is ~even}}}\prod_{i\in \mathcal{A}}\beta_i\prod_{j\in \mathcal{I}\setminus\mathcal{A}}\alpha_j.\\
\beta_{\rm out}&=\sum_{\substack{\mathcal{A}\subseteq \mathcal{I}\\|\mathcal{A}|{\rm ~is ~odd}}}\prod_{i\in \mathcal{A}}\beta_i\prod_{j\in  \mathcal{I}\setminus\mathcal{A}}\alpha_j.\\
\gamma_{\rm out}&=1-\prod_{j=0}^m(1-\gamma_j).
\end{align}
\item[2)] {\bf Node} \fbox{=}: From \fbox{=} to \fbox{C}, given the input messages $\{(\alpha_{\ell},\beta_{\ell}, \gamma_{\ell}),{\ell}\in  \mathcal{I}\}$ from the other edges, the extrinsic output messages can be calculated as
\begin{align}
\alpha_{\rm out}&= \sum_{\substack{\mathcal{A},\mathcal{B}\subseteq \mathcal{I}\\ \mathcal{A}\cap\mathcal{B}=\varnothing\\|\mathcal{A}|>|\mathcal{B}|}}\prod_{i\in \mathcal{A}}\alpha_i\prod_{j\in \mathcal{B}}\beta_j\!\!\!\!\!\prod_{\ell\in  \mathcal{I}\setminus (\mathcal{A}\cup\mathcal{B})}\!\!\!\!\!\gamma_\ell,\\
\beta_{\rm out}&= \sum_{\substack{\mathcal{A},\mathcal{B}\subseteq \mathcal{I}\\ \mathcal{A}\cap\mathcal{B}=\varnothing\\|\mathcal{A}|<|\mathcal{B}|}}\prod_{i\in \mathcal{A}}\alpha_i\prod_{j\in \mathcal{B}}\beta_j\!\!\!\!\!\prod_{\ell\in  \mathcal{I}\setminus (\mathcal{A}\cup\mathcal{B})}\!\!\!\!\!\gamma_\ell,\\
\gamma_{\rm out}&=\sum_{\substack{\mathcal{A},\mathcal{B}\subseteq \mathcal{I}\\ \mathcal{A}\cap\mathcal{B}=\varnothing\\|\mathcal{A}|=|\mathcal{B}|}}\prod_{i\in \mathcal{A}}\alpha_i\prod_{j\in \mathcal{B}}\beta_j\!\!\!\!\!\prod_{\ell\in  \mathcal{I}\setminus (\mathcal{A}\cup\mathcal{B})}\!\!\!\!\!\gamma_\ell.
\end{align}
From \fbox{=} to \fbox{+}, given the input messages $\{(\alpha_{\ell},\beta_{\ell}, \gamma_{\ell}),{\ell} \in \mathcal{I}\}$ from the other edges (label the input message from node \fbox{C} with index $0$), the output messages are given by
\begin{align}
\alpha_{\rm out}=&\alpha_0+\gamma_0\!\!\!\!\!\!\sum_{\substack{\mathcal{A},\mathcal{B}\subseteq \mathcal{I}\setminus \{0\}\\ \mathcal{A}\cap\mathcal{B}=\varnothing\\|\mathcal{A}|>|\mathcal{B}|}}\prod_{i\in \mathcal{A}}\alpha_i\prod_{j\in \mathcal{B}}\beta_q\!\!\!\!\!\prod_{\ell\in \mathcal{I}\setminus (\{0\}\cup\mathcal{A}\cup\mathcal{B})}\!\!\!\!\!\gamma_\ell,\\
\beta_{\rm out}=&\beta_0+\gamma_0\!\!\!\!\!\!\sum_{\substack{\mathcal{A},\mathcal{B}\subseteq\mathcal{I}\setminus \{0\}\\ \mathcal{A}\cap\mathcal{B}=\varnothing\\|\mathcal{A}|<|\mathcal{B}|}}\prod_{i\in \mathcal{A}}\alpha_i\prod_{j\in \mathcal{B}}\beta_q\!\!\!\!\!\prod_{\ell\in\mathcal{I}\setminus (\{0\}\cup\mathcal{A}\cup\mathcal{B})}\!\!\!\!\!\gamma_\ell,\\
\gamma_{\rm out}=&\gamma_0\!\!\!\!\!\!\sum_{\substack{\mathcal{A},\mathcal{B}\subseteq\mathcal{I}\setminus \{0\}\\ \mathcal{A}\cap\mathcal{B}=\varnothing\\|\mathcal{A}|=|\mathcal{B}|}}\prod_{i\in \mathcal{A}}\alpha_i\prod_{j\in \mathcal{B}}\beta_q\!\!\!\!\!\prod_{\ell\in\mathcal{I}\setminus (\{0\}\cup\mathcal{A}\cup\mathcal{B})}\!\!\!\!\!\gamma_\ell.
\end{align}
\item[3)]{\bf Node} \fbox{C}: Given the input messages $(\alpha,\beta,\gamma)$ from node \fbox{=}, the output messages can be calculated as
\begin{align}\label{equ:DEBER0}
\!\!\!\!\!\!\alpha_{\rm out}&=\!\sum_{i,j}(1-\mu_{i,j}-\lambda_{i,j})f_{i,j}(\alpha,\beta,\gamma),\\
\!\!\!\!\!\!\beta_{\rm out}&=\!\sum_{i,j}\mu_{i,j}f_{i,j}(\alpha,\beta,\gamma),\\
\!\!\!\!\!\!\gamma_{\rm out}&=\!\sum_{i,j}\lambda_{i,j}f_{i,j}(\alpha,\beta,\gamma),\\
\!\!\!\!\!\!{\rm BER} &=\!\sum_{i,j}[\mu_{i,j}+\lambda_{i,j}(\frac{i}{n}+\frac{j}{2n})]f_{i,j}(\alpha,\beta,\gamma).
\label{equ:DEBER1}
\end{align}
\end{itemize}
%
%
%
%




\bibliographystyle{IEEEtran}
\end{document}